\documentstyle[aps,twocolumn]{revtex}

\begin{document}
\title{Thermodynamics of isotropic and anisotropic layered magnets: renormalization
group approach and $1/N$ expansion}
\author{V.Yu.Irkhin and A.A.Katanin$^{*}$}
\address{620219, Institute of Metal Physics, Ekaterinburg, Russia.}
\maketitle

\begin{abstract}
The $O(N)$ model of layered antiferro- and ferromagnets with a weak
interlayer coupling and/or easy-axis anisotropy is considered. A
renormalization group (RG) analysis in this model is performed, the results
for $N=3$ being expected to agree with those of the $1/M$ expansion in the $%
CP^{M-1}$model at $M=2$. The quantum and classical cases are considered. A
crossover from an isotropic 2D-like to 3D Heisenberg (or 2D Ising) regime is
investigated within the $1/N$ expansion. Analytical results for the
temperature dependence of the (sublattice) magnetization are obtained in
different regimes. The RG results for the ordering temperature are derived.
In the quantum case they coincide with the corresponding results of the $1/N$
expansion. The numerical calculations on the base of the equations obtained
yield a good agreement with experimental data on the layered perovskites La$%
_2$CuO$_4,$ K$_2$NiF$_4$ and Rb$_2$NiF$_4,$ and the Monte-Carlo results for
the anisotropic classical systems.
\end{abstract}


\section{Introduction}

The problem of layered magnetic systems is of interest both from theoretical
and practical point of view. Here belong, e.g., quasi-two-dimensional
(quasi-2D) perovskites\cite{Joungh}, ferromagnetic monolayers and ultrathin
films\cite{Elmers}. Such systems possess magnetic transition temperatures
which are low in comparison with the intralayer exchange parameter $J$ and
are determined by magnetic anisotropy and/or interlayer coupling.

The crucial role in the systems with small interlayer coupling (or
anisotropy) belongs to the temperature crossover from a ``2D-like''
(isotropic) regime to the critical 3D (or 2D Ising) regime respectively
(see, e.g., Refs.\cite{Joungh,Quasi2D}). In the general case where both
interlayer coupling and anisotropy parameter are present, the situation is
more complicated: with increasing temperature the 2D-like Heisenberg
behavior changes to the 2D-Ising or 3D-Heisenberg one depending on that
anisotropy or interlayer coupling dominate, and finally system passes to the
3D Ising behavior.

There exist a number of approximations which treat thermodynamics of layered
systems. The standard spin-wave theory (SWT) describes satisfactorily the
region of rather low temperatures only. Somewhat better results can be
obtained by taking into account the temperature renormalization of the
interlayer coupling parameter and the anisotropy parameter. The temperature
dependence of the anisotropy parameter within the spin-wave theory was
considered in Refs.\cite{Oguchi,Rastelli,Wang}. More systematic way to
consider such renormalizations is the self-consistent spin-wave theory (SSWT)%
\cite{Takahashi,ArovasBook} which was applied to quasi-2D and anisotropic 2D
magnets in Refs.\cite{Our1st,Liu,OurFMM,OurSSWT}.

SSWT takes into account the interaction between spin waves in the lowest
Born approximation. However, at not too low temperatures this approximation
is insufficient. In particular, the values of the ordering temperature in
SSWT are still too high in comparison with experimental ones, and the
critical behavior is quite incorrect. Thus the summation of leading
contributions in all orders of perturbation theory should be performed. At
the same time, to describe the behavior of the order parameter in the
critical region one has to take into consideration fluctuation
(non-spin-wave) contribution to thermodynamic quantities. Ising-like
excitations in a classical anisotropic model were considered in Ref.\cite
{Levanjuk}. At the same time, in the quantum anisotropic case this treatment
meets with difficulties\cite{Levanjuk2}, and 3D fluctuations in the critical
region of quasi-2D magnets cannot be considered in the approach \cite
{Levanjuk,Levanjuk2} too.

A possible way to sum up an infinite sequence of perturbation contributions
is the renormalization group (RG) analysis. The RG approach was successfully
used to consider the classical isotropic magnets with the space
dimensionality $d=2+\varepsilon $ \cite{Brezin,Pelkovitz}. In this case the
fixed point is small, $T^{*}\sim \varepsilon J$ $(J$ is the exchange
parameter), and the standard technique of the $\varepsilon $-expansion can
be applied. Physically, this means that the excitation-spectrum picture
differs somewhat from the spin-wave one (as discussed in Ref.\cite{Quasi2D},
the fluctuation corrections to the excitation spectrum read $\delta E_{{\bf q%
}}\sim \varepsilon \ln q$). The RG method was applied also to quantum 2D
isotropic magnets in Ref.\cite{Chakraverty}.

The scaling behavior in the quasi-2D or anisotropic 2D systems is expected
to differ from the isotropic $d=2+\varepsilon $ magnet. In these cases the
renormalized value of $T^{*}/J$ at the stability point of the RG
transformation is not small, which corresponds to above-discussed crossover
from the 2D-like Heisenberg to 3D Heisenberg (or 2D Ising) critical regime.
Thus the RG method does not work when passing to the true critical region.
The latter region should to be considered with account of essentially
non-spin-wave (fluctuation) excitations.

To take into account non-spin-wave (fluctuation) effects it is convenient to
use, instead of the original Heisenberg model, models with large degeneracy,
which enables one to introduce a formall small parameter in the theory. In
Ref.\cite{Quasi2D} an isotropic quantum quasi-2D antiferromagnet was
investigated within the $1/N$ expansion in the quantum $O(N)$ model\cite
{Chubukov} (in the Heisenberg model, $N=3$). It was demonstrated that the
renormalizations of the interlayer coupling parameter in comparison with the
usual spin-wave theory (or SSWT) determine considerable lowering of the
transition point. The same situation should be expected for 2D anisotropic
magnets. At the same time, the $1/N$ expansion meets with some difficulties
at the description of the 2D-like fluctuations in the renormalized-classical
temperature region near $d=2$ where the series expansion is performed in
powers of $1/(N-2)$ rather than $1/N$ \cite{Chubukov,Quasi2D}.

Thus the RG approach and $1/N$ expansion in the $O(N)$ model are expected to
have advantages in different temperature regions. Whereas the first method
describes well the 2D-like regime, the $1/N$ expansion treats satisfactorily
the critical region. An advantage of the RG approach in comparison with the
technique of the $1/N$ expansion in the $O(N)$ model is that it enables one
to consider the quantum ferromagnet case where the partition function cannot
be generalized to arbitrary $N.$ The RG analysis permits also to treat the
classical case, which is difficult within the $1/N$ expansion. (In the
classical case, there is no natural upper cutoff for quasimomenta, which is
the temperature in the quantum case, and the original lattice version of the
partition function should be considered.)

The Heisenberg model can be also considered as a particular case of the $%
SU(M)$ model (or of its continual analog, $CP^{M-1}$ model) with $M=2$ (see,
e.g., \cite{Read}). Since the $M\rightarrow \infty $ limit corresponds to
SSWT (see, e.g., Ref.\cite{ArovasBook}), at finite $M$ thermodynamics is
described in terms of spin-wave picture of the excitation spectrum. The
corresponding $1/M$-expansion contains in the 2D case infrared-divergent
terms \cite{Starykh} and is rather inconvenient. For $d$ not close to $2$,
this expansion does not enable one to obtain correct values of the critical
exponents\cite{OurCP}. Thus, as well as in RG approach, only the 2D-like
isotropic Heisenberg region can be considered within the $1/M$-expansion.
Furthermore, the results of $CP^{M-1}$ model to $n$-th order in $1/M$ are
expected to coincide at $M=2$ with the $(n+1)$-loop RG analysis for $O(3)$
or, equivalently, $CP^1$ model (we do not know a general proof of this
statement, but this is true in the $d=2+\varepsilon $ case for $n=0,1,$ see
Ref.\cite{OurCP}).

The simplest one-loop RG analysis was applied earlier to calculation of the
Curie temperature of anisotropic 2D ferromagnets \cite{RGA}. As we shall see
below, the results of this work become considerably modified by the two-loop
corrections which were not taken into account in Ref.\cite{RGA}. The
momentum-shell version of the one-loop RG approach in the classical quasi-2D
magnets was considered in \cite{RGQ}. However, the authors of this paper
passed to continual limit in the direction perpendicular to layers, so that
results of Ref.\cite{RGQ} at small interlayer coupling have qualitative
rather quantitative character.

In the present paper we consider thermodynamics of the quantum and classical
layered magnets with small interlayer coupling and anisotropy within the
consistent two-loop RG approach and first-order $1/N$ expansion in the $O(N)$
model.

The plan of the paper is as follows. In Sect. 2 we consider the continual
and lattice versions of the $O(N)$ model for anisotropic layered ferro- and
antiferromagnets. In Sect. 3 we apply the renormalization group approach
\cite{Brezin,Chakraverty} to the quantum $O(N)$ model in the (isotropic)
2D-like region (the classical case is considered in Appendix B). In Sect. 4
we investigate the same problem within the $1/N$ expansion in the $O(N)$
model up to the first-order and treat the crossover from the 2D-like to 3D
Heisenberg (or 2D Ising) behavior. In Sect.5 we summarize our results and
compare them with experimental data on layered antiferromagnets.

\section{Continual and lattice models for the spin system}

We consider the Heisenberg model with small interlayer coupling and
easy-axis anisotropy
\begin{equation}
H=-\frac J2\sum_{i\delta _{\parallel }}{\bf S}_i{\bf S}_{i+\delta
_{\parallel }}+H_{\text{3D}}+H_{\text{anis}}  \label{H}
\end{equation}
\begin{eqnarray}
\,H_{\text{3D}} &=&-\frac{\alpha J}4\sum_{i\delta _{\perp }}{\bf S}_i{\bf S}%
_{i+\delta _{\perp }}, \\
\,\,\,\,\,\,H_{\text{anis}} &=&-\frac{J\eta }2\sum_{i\delta _{\parallel
}}S_i^zS_{i+\delta _{\parallel }}^z-|\,J\,|\zeta \sum_i(S_i^z)^2
\end{eqnarray}
where $J>0$ for a ferromagnet, $J<0$ for an antiferromagnet, $\delta
_{\parallel }$ and $\delta _{\perp }$ denote nearest neighbors within a
layer and for different layers,$\,\,\alpha >0$ is the interlayer coupling
parameter, $\eta ,\zeta >0$ are the parameters of the exchange anisotropy
and single-site anisotropy respectively. The partition function of the model
(\ref{H}) can be represented in terms of a path integral over coherent
states (see, e.g., Refs.\cite{Klauder,ArovasBook}):
\begin{eqnarray}
{\cal Z} &=&\int D{\bf n}D\lambda \exp \left\{ \frac{JS^2}2%
\int\limits_0^{1/T}d\tau \sum_i\left[ \frac{2\,\text{i}}{JS}{\bf A}({\bf n}%
_i)\frac{\partial {\bf n}_i}{\partial \tau }+{\bf n}_i{\bf n}_{i+\delta
_{\parallel }}\right. \right.  \label{zp} \\
&&\left. \left. +\frac \alpha 2{\bf n}_i{\bf n}_{i+\delta _{\perp }}+\eta
n_i^zn_{i+\delta _{\parallel }}^z+\,\text{sgn}(J)\widetilde{\zeta }%
(n_i^z)^2+hn_i^z+\text{i\thinspace }\lambda _i({\bf n}_i^2-1)\right] \right\}
\nonumber
\end{eqnarray}
with ${\bf n}_i(\tau )$ the three-component unit-length vector field, ${\bf A%
}({\bf n})$ the vector potential of the unit magnetic monopole, which
satisfies the equation ${\bf \nabla }\times {\bf A(n)}\cdot {\bf n}=1,$ $%
\widetilde{\zeta }=2\zeta (1-1/2S)$ and the summation over $\delta _{\Vert
},\delta _{\perp }$ in (\ref{zp}) is assumed. We have also introduced in (%
\ref{zp}) the external magnetic field $h$ to perform the calculation of spin
correlation functions. The term with the time derivative corresponds to the
Berry-phase contribution \cite{Haldane}. Depending on the value of $T/JS,$
two cases are possible: (a) the classical case $T\gg JS$ and (b) the quantum
case $T\ll JS$.

Consider first the classical case. The main contribution to (\ref{zp}) comes
from time-independent paths, and the partition function reduces to 
\begin{eqnarray}
{\cal Z}_{\text{cl}} &=&\int D{\bf n}D\lambda \exp \left\{ \frac{\rho _s^0}{%
2T}\sum_i\left[ {\bf n}_i{\bf n}_{i+\delta _{\parallel }}+\frac \alpha 2{\bf %
n}_i{\bf n}_{i+\delta _{\perp }}\right. \right.  \label{zcl} \\
&&\ \left. \left. +\eta n_i^zn_{i+\delta _{\parallel }}^z+\widetilde{\zeta }%
(n_i^z)^2+hn_i^z+\text{i\thinspace }\lambda ({\bf n}_i^2-1)\right] \right\} 
\nonumber
\end{eqnarray}
with $\rho _s^0=|\,J\,|\,S^2$ the bare spin stiffness. To derive (\ref{zcl})
in the antiferromagnetic case we have replaced ${\bf n}_i\rightarrow -{\bf n}%
_i,\,\lambda _i\rightarrow -\lambda _i$ at one of two sublattices. Thus in
the classical case the results for ${\cal Z}$ are identical for a ferro- and
antiferromagnet. In the continual limit the partition function (\ref{zcl})
coincides with that of the well-known classical nonlinear-sigma model \cite
{ArovasBook}. However, if one is interested in thermodynamics in a wide
temperature interval (not only in the critical region), the continual limit
cannot be used since not only long-wave excitations contribute thermodynamic
properties.

Now we treat the quantum case. Then the temperature plays the role of a
natural upper limit cutoff for frequencies of the fluctuations. Thus we may
pass to the continual limit within each layer. For a ferromagnet we use the
representation 
\begin{equation}
{\bf A}({\bf n})=\frac{{\bf z\times n}}{1+({\bf zn)}}
\end{equation}
(${\bf z}$ is the unit vector along the $z$-axis)${\bf .}$ Then we obtain 
\begin{eqnarray}
{\cal Z}_{\text{F}} &=&\int \frac{D\mbox {\boldmath $\pi $}}{\sqrt{1-\pi ^2}}%
\exp \left\{ -\frac{\rho _s^0}2\int\limits_0^{1/T}d\tau \int d^2{\bf r}%
\sum\limits_{i_z}\left[ \frac{2\text{i}}{JS}\text{\thinspace }\frac{1-\sqrt{%
1-\pi _{i_z}^2}}{\pi _{i_z}^2}\right. \right.  \nonumber \\
&&\times \left( \pi _{i_z}^x\frac{\partial \pi _{i_z}^y}{\partial \tau }-\pi
_{i_z}^y\frac{\partial \pi _{i_z}^x}{\partial \tau }\right) +(\nabla 
\mbox
{\boldmath $\pi $}_{i_z})^2+\frac \alpha 2(\mbox {\boldmath $\pi $}_{i_z+1}-%
\mbox {\boldmath $\pi $}_{i_z})^2  \nonumber \\
&&+(\nabla \sqrt{1-\mbox {\boldmath $\pi $}_{i_z}^2})^2+\frac \alpha 2(\sqrt{%
1-\mbox {\boldmath $\pi $}_{i_z+1}^2}-\sqrt{1-\mbox {\boldmath $\pi $}%
_{i_z}^2})^2+f\mbox {\boldmath $\pi $}_{i_z}^2  \nonumber \\
&&\left. \left. +h\sqrt{1-\mbox {\boldmath $\pi $}_{i_z}^2}\right] \right\}
\label{zf}
\end{eqnarray}
where $i_z$ is the number of a layer,$\,\mbox {\boldmath $\pi $}={\bf n-(nz)z%
}$ is the two-component vector field, 
\begin{equation}
f=\widetilde{\zeta }+4\eta \equiv 2\zeta (1-1/2S)+4\eta
\end{equation}
and we have made the shift $i\lambda \rightarrow i\lambda +f$ before
integrating over $\lambda .$

In the antiferromagnetic quantum case we use the Haldane mapping \cite
{Haldane} (see also Ref.\cite{ArovasBook}) to integrate over the ``fast''
components of ${\bf n.}$ Thus we pass to the partition function of the
quantum nonlinear sigma model 
\begin{eqnarray}
{\cal Z}_{\text{AF}} &=&\int D\mbox {\boldmath $\sigma $}D\lambda \exp
\left\{ -\frac{\rho _s^0}2\int\limits_0^{1/T}d\tau \int d^2{\bf r}%
\sum_{i_z}\left[ \frac 1{c_0^2}(\partial _\tau \mbox {\boldmath $\sigma $}%
_{i_z})^2\right. \right.  \nonumber \\
&&\ +({\bf \nabla }\mbox {\boldmath $\sigma $}_{i_z})^2+\frac \alpha 2(%
\mbox
{\boldmath $\sigma $}_{i_z+1}-\mbox {\boldmath $\sigma $}_{i_z})^2-f(\sigma
_{i_z}^z)^2+h\sigma _{i_z}^z  \label{za} \\
&&\ \left. \left. +i\lambda (\mbox {\boldmath $\sigma $}_{i_z}^2-1)\right]
\right\}
\end{eqnarray}
where $\mbox {\boldmath $\sigma $}_{i_z}$ is the three-component unit-length
field and $c_0=\sqrt{8}JS$ is the bare spin-wave velocity. Unlike the
quantum ferromagnet case, this model can be trivially extended to the $O(N)$
symmetry with arbitrary $N$, the $N$-component vector field $\sigma
_i=\{\sigma _1...\sigma _N\}$ being introduced and $\sigma ^z$ being
replaced by $\sigma _N$. Writing down $\sigma =\{\pi _1...\pi _{N-1},\sigma
_N\}$ we have 
\begin{eqnarray}
{\cal Z}_{\text{AF}} &=&\int \frac{D\pi }{\sqrt{1-\pi ^2}}\exp \left\{ -%
\frac{\rho _s^0}2\int\limits_0^{1/T}d\tau \int d^2{\bf r}\sum\limits_{i_z}%
\left[ (\partial _\mu \pi _{i_z})^2\right. \right.  \nonumber \\
&&\ \ +(\partial _\mu \sqrt{1-\pi _{i_z}^2})^2+\frac \alpha 2(\pi
_{i_z+1}-\pi _{i_z})^2+f\pi ^2  \nonumber \\
&&\ \ \left. \left. +\frac \alpha 2(\sqrt{1-\pi _{i_z+1}^2}-\sqrt{1-\pi
_{i_z}^2})^2+h\sqrt{1-\pi _{i_z}^2}\right] \right\}  \label{zaf}
\end{eqnarray}
where $\partial _\mu =(\partial /\partial (c_0\tau ),\nabla )$, and we have
performed the shift $i\lambda \rightarrow i\lambda +f.$

\section{Renormalization group analysis in the 2D-like regime for the
quantum case}

Using the above expressions for the partition function we can develop a
scaling approach. We use the field theory formulation of RG transformation 
\cite{Amit,Brezin}. To develop this approach we pass to the renormalized
theory with the use of the relations \cite{Brezin} 
\begin{eqnarray}
g &=&g_RZ_1,\,u=u_RZ_u  \nonumber \\
\,\pi &=&\pi _RZ,\,\,h=h_RZ_1/\sqrt{Z}  \nonumber \\
\,f &=&f_RZ_2,\,\alpha =\alpha _RZ_3  \label{ZR}
\end{eqnarray}
where the index $R$ corresponds to the renormalized quantities, the bare
coupling constant $g$ and the dimensionless inverse temperature $u$ are
determined by 
\begin{eqnarray}
g &=&1/S,\,\,u=JS/T\;\;\,\,\,\,\,\,\text{(FM)} \\
g &=&c_0/\rho _s^0,\;u=c_0/T\;\text{(AFM),}
\end{eqnarray}
The renormalization constants $Z_i$ are chosen from the condition that the
thermodynamic quantities are independent of an upper cutoff. Since the
nonlinear-sigma model is renormalizable (see Ref. \cite{Amit}), five
renormalization constants for five independent parameters of the model are
sufficient to this end. To calculate renormalization constants it is
sufficient to calculate the renormalization of the one-particle Green's
function in an external magnetic field \cite{Amit,Brezin}.

Consider first the case of an antiferromagnet. The perturbation theory in
the coupling constant for the partition function (\ref{zaf}) can be
developed in a standard way\cite{Brezin,Pelkovitz,Chakraverty}. To this end
we expand the square roots in a series in the field $\pi $. The bare Green's
function of this field has the form 
\begin{equation}
G^{(0)}({\bf p},i\omega _n)=\frac 1g[\omega _n^2+p_{\parallel }^2+\alpha
(1-\cos p_z)+f+h]^{-1}  \label{G0AF}
\end{equation}
where $p_{\parallel }=(p_x^2+p_y^2)^{1/2},$ $\omega _n=2\pi n/u$ are the
dimensionless Matsubara frequencies. In each order of perturbation theory
one has to take into account all the possible connected diagrams with the
fixed number of loops.

We consider only the ``renormalized classical'' regime 
\begin{equation}
T\gg (\max \{f,\alpha \})^{1/2}c  \label{cond}
\end{equation}
since in the interval $T<(\max \{f,\alpha \})^{1/2}c$ the staggered
magnetization is well described by SSWT (see a more detailed discussion of
different temperature regimes in Ref.\cite{Quasi2D}). Since in this regime
the quasimomentum cutoff parameter for quantum fluctuations is the boundary
of the Brillouin zone $\Lambda ,$ while for thermal ones this is $T/c,$
effects of quantum and thermal fluctuations can be separated (see e.g. \cite
{Chubukov,Quasi2D}). Thus it is useful to perform the renormalization in two
steps. At the first step the ground-state quantum renormalizations are
performed, and the second step is the temperature renormalization. To this
end we represent the renormalization constants as 
\begin{equation}
Z_i(g,u,\Lambda )=Z_{Qi}(g,\Lambda )\widetilde{Z}_i(g_r,u_r)  \label{ZZ}
\end{equation}
where $Z_{Qi}(g)$ contain ground-state quantum renormalizatons, and $%
\widetilde{Z}_i$ all the others, 
\begin{equation}
g_r=Z_{Q1}^{-1}g,\,\;u_r=Z_{Qu}^{-1}u  \label{grur}
\end{equation}
are the quantum-renormalized coupling constant and dimensionless inverse
temperature. Note that the latter is renormalized due to the renormalization
of the spin-wave velocity, $c=Z_{Qu}c_0.$ For further convenience we also
introduce the quantum-renormalized anisotropy and interlayer coupling
parameters 
\begin{equation}
f_r=Z_{Q2}^{-1}f,\,\,\,\alpha _r=Z_{Q3}^{-1}\alpha  \label{QRP}
\end{equation}
which are just experimentally observed. Up to one-loop order we have 
\begin{eqnarray}
Z_Q &=&1-(N-1)\frac{g\Lambda }{4\pi }+{\cal O}(g^2),\,\,  \nonumber \\
\,Z_{Q1} &=&1-(N-2)\frac{g\Lambda }{4\pi }+{\cal O}(g^2),\,\,Z_{Qu}=1+{\cal O%
}(g^2)  \nonumber \\
Z_{Q2} &=&1+\frac{g\Lambda }{2\pi }+{\cal O}(g^2),\,\,Z_{Q3}=1+\frac{%
3g\Lambda }{4\pi }+{\cal O}(g^2)  \label{ZQ}
\end{eqnarray}
Since the renormalization constants $Z_{Qi}$ are non-universal, the
continual limit is insufficient to calculate them, and quantum-renormalized
parameters can be determined only from the consideration of the original
lattice partition function (\ref{zp}) in the ground state. For the
square-lattice antiferromagnet this can be performed within the spin-wave
theory, which is in fact a series expansion in $g$ ($g\sim 1/S$ for large $S$%
). The results of the spin-wave theory are presented in Appendix A. We have
to first-order in $1/S$%
\begin{eqnarray}
Z_Q &=&1/Z_{Q1}=Z_{Q2}=Z_{Q3}^{1/2}=1-0.197/S,  \label{ZQSW} \\
\,\,Z_{Qu} &=&1+0.079/S.  \nonumber
\end{eqnarray}
Note that after performing the ground-state quantum renormalizations (\ref
{ZQ}) in the continual model (or, equivalently, renormalizations (\ref{ZQSW}%
) in the original lattice model) the theory becomes completely universal,
i.e. thermodynamic properties do not depend on the cutoff parameter $\Lambda
.$

Now we pass to the consideration of the temperature renormalizations. The
calculation of the renormalization constants $\widetilde{Z}_i$ is performed
in the same way as in Ref.\cite{Brezin} and leads to $\widetilde{Z}_u\equiv
1,$ 
\begin{eqnarray}
\widetilde{Z} &=&1+t_r(N-1)\ln (u_r\mu )  \nonumber \\
&&+t_r^2(N-1)(N-3/2)\ln ^2(u_r\mu )+{\cal O}(t_r^3),  \nonumber \\
\widetilde{Z}_1 &=&1+t_r(N-2)\ln (u_r\mu )+t_r^2(N-2)\ln ^2(u_r\mu )+{\cal O}%
(t_r^3),\,  \nonumber \\
\widetilde{Z}_2 &=&1-2t_r\ln (u_r\mu )+{\cal O}(t_r^2),\,\,\,\,\widetilde{Z}%
_3=1-t_r\ln (u_r\mu )+{\cal O}(t_r^2),  \label{RAF}
\end{eqnarray}
$\,$where $\mu $ is an infrared cutoff parameter with the dimensionality of
the inverse length and $t_r=g_r/(2\pi u_r)$. The only difference from the
results of Ref.\cite{Brezin} is that the ultraviolet cutoff parameter in (%
\ref{RAF}) is $u_r$ (rather than $\Lambda $ in the classical case), and two
new renormalization constants, $\widetilde{Z}_2$ and $\widetilde{Z}_3,$ for
the anisotropy and interlayer coupling parameters occur.

The infinitesimal change of $\mu $ generates the renormalization group
transformation, and the derivatives of $Z$-factors with respect to $\mu $
determine the renormalized-parameters flow functions (see, e.g., Ref.\cite
{Amit}). Since quantum-renormalization constants (\ref{ZQ}) are invariant
under RG transformation, it is sufficient to calculate the derivatives of $%
\widetilde{Z}.$ To the two-loop order we have 
\begin{eqnarray}
\beta (t_r) &\equiv &\mu \frac{dt_r}{d\mu }=-(N-2)t_r^2-(N-2)t_r^3+{\cal O}%
(t_r^4)  \label{f1} \\
\varsigma (t_r) &\equiv &\mu \frac{d\ln Z}{d\mu }=(N-1)t_r+{\cal O}(t_r^3)
\label{f2}
\end{eqnarray}
Other two flow functions will be needed only to the one-loop approximation: 
\begin{eqnarray}
\gamma _f(t_r) &\equiv &\mu \frac{d\ln Z_2}{d\mu }=-2t_r+{\cal O}(t_r^2)
\label{f3} \\
\gamma _\alpha (t_r) &\equiv &\mu \frac{d\ln Z_3}{d\mu }=-t_r+{\cal O}(t_r^2)
\label{f4}
\end{eqnarray}
Using (\ref{f1})-(\ref{f4}) we find the scaling laws for the Hamiltonian
parameters under RG transformation 
\begin{eqnarray}
\rho &=&\exp \left[ \,\,\int\limits_{t_r}^{t_\rho }\frac{dt}{\beta (t)}%
\right]  \nonumber \\
\ &=&\left( \frac{t_\rho }{t_r}\right) ^{1/(N-2)}\exp \left[ \frac 1{N-2}%
\left( \frac 1{t_\rho }-\frac 1{t_r}\right) \right] \left[ 1+{\cal O}(t_\rho
)\right] ,  \label{r1} \\
Z_\rho &=&\exp \left[ -\int\limits_{t_r}^{t_\rho }\frac{\varsigma (t)}{\beta
(t)}dt\right]  \nonumber \\
\ &=&\left( \frac{t_\rho }{t_r}\right) ^{(N-1)/(N-2)}\left[ 1+\frac{N-1}{N-2}%
(t_r-t_\rho )+{\cal O}(t_\rho ^2)\right] ,  \label{r2} \\
f_\rho &=&f_r\exp \left[ -\int\limits_{t_r}^{t_\rho }\frac{\gamma _f(t)}{%
\beta (t)}dt\right] =f_r\left( \frac{t_\rho }{t_r}\right) ^{-2/(N-2)}\left[
1+{\cal O}(t_\rho )\right] ,  \label{r3} \\
\alpha _\rho &=&\alpha _r\exp \left[ -\int\limits_{t_r}^{t_\rho }\frac{%
\gamma _\alpha (t)}{\beta (t)}dt\right] =\alpha _r\left( \frac{t_\rho }{t_r}%
\right) ^{-1/(N-2)}\left[ 1+{\cal O}(t_\rho )\right] .  \label{r4}
\end{eqnarray}
where $t_\rho ,f_\rho ,\alpha _\rho $ are the corresponding parameters of
the effective Hamiltonian at the scale $\mu \rho ,$ and $Z_\rho $ is the
scaling factor for the spin fields on this scale.

Now we are ready to calculate the relative sublattice magnetization $%
\overline{\sigma }_r=\overline{\sigma }/\overline{\sigma }_0$, where $%
\overline{\sigma }=\overline{S}/S$ ($\overline{S}=\langle S_{{\bf Q}%
}^z\rangle $ is the staggered magnetization) and $\overline{\sigma }_0\equiv 
\overline{\sigma }(T=0)$. The perturbation result in the zero magnetic field
up to terms of the order of $t_r^2$ reads 
\begin{eqnarray}
\overline{\sigma }_r &=&1-\frac{t_r(N-1)}4\ln \frac 2{u_r^2\Delta
(f_r,\alpha _r)}  \nonumber \\
&&\ \ +\frac{t_r^2(3-N)(N-1)}{32}\ln ^2\frac 2{u_r^2\Delta (f_r,\alpha _r)} 
\nonumber \\
&&\ \ -\frac{t_r^2(N-1)(B_2-2)}8\ln \frac 2{u_r^2\Delta (f_r,\alpha _r)}
\label{MPAF}
\end{eqnarray}
where 
\begin{eqnarray}
\Delta (f,\alpha ) &=&f+\alpha +\sqrt{f^2+2\alpha f}  \label{Delta} \\
B_2 &=&3+f_r/\sqrt{f_r^2+2\alpha _rf_r}  \label{b2}
\end{eqnarray}
Note that the last term in (\ref{MPAF}) corresponds to temperature
renormalizations of the interlayer coupling and anisotropy parameters: 
\begin{eqnarray}
f_t &=&f_r\left[ 1-t_r\ln \frac 2{u_r^2\Delta (f_r,\alpha _r)}\right] 
\label{fr} \\
\alpha _t &=&\alpha _r\left[ 1-\frac{t_r}2\ln \frac 2{u_r^2\Delta
(f_r,\alpha _r)}\right]   \label{ar}
\end{eqnarray}
We obtain from (\ref{r2})\ the following scaling law for the sublattice
magnetization 
\begin{equation}
\overline{\sigma }_r(t_r,f_r,\alpha _r,\mu _r)=Z_\rho ^{-1/2}\overline{%
\sigma }_r(t_\rho ,f_\rho ,\alpha _\rho ,\mu \rho )
\end{equation}
or, equivalently, 
\begin{equation}
\frac{\overline{\sigma }_r(t_r,f_r,\alpha _r,\mu )}{\overline{\sigma }%
_r(t_\rho ,f_\rho ,\alpha _\rho ,\mu \rho )}=\left( \frac{t_r}{t_\rho }%
\right) ^{\beta _2}\left[ 1-\beta _2(t_r-t_\rho )+{\cal O}(t_\rho ^2)\right] 
\label{mrg}
\end{equation}
where 
\begin{equation}
\beta _2=\frac{N-1}{2(N-2)}
\end{equation}
is the sublattice-magnetization ``critical exponent'' in the temperature
interval under consideration. This coincides with the $\varepsilon
\rightarrow 0$ limit of the critical exponent $\beta _{2+\varepsilon }$ for $%
d=2+\varepsilon $ \cite{Brezin}$.$ The equation for $t_\rho /t_r$ is given
by (\ref{r1}) which can be rewritten as 
\begin{equation}
\frac{t_r}{t_\rho }=1+t_r\ln (\frac{t_r}{t_\rho }\rho ^{N-2})+t_r{\cal O}%
(t_\rho )  \label{trg}
\end{equation}
Finally, the scale $\rho $ is fixed by the condition that the arguments of
logarithms in (\ref{MPAF}) on this scale are equal to unity, i.e. by $%
\overline{\sigma }_r(t_\rho ,f_\rho ,\alpha _\rho ,\mu \rho )=1.$ Taking
into account that $u$ scales in a trivial way, $u_\rho =u_r/\rho ,$ we
obtain the additional equation for $\rho $: 
\begin{equation}
2\rho ^2=u_r^2\Delta (f_\rho ,\alpha _\rho )  \label{eq}
\end{equation}
We have from (\ref{mrg}), (\ref{trg}), (\ref{r3}), (\ref{r4}) and (\ref{eq})
the equation for the relative sublattice magnetization in the two-loop RG\
analysis 
\begin{eqnarray}
\overline{\sigma }_r^{1/\beta _2} &=&1-\frac{t_r}2\left[ (N-2)\ln \frac 2{%
u_r^2\Delta (f_t,\alpha _t)}+\frac 2{\beta _2}\ln (1/\overline{\sigma }%
_r)\right.   \label{MRAF} \\
&&\ +2(1-\overline{\sigma }_r^{1/\beta _2})+{\cal O}(t_r/\overline{\sigma }%
_r^{1/\beta _2})\bigg]  \nonumber
\end{eqnarray}
where $f_t$ and $\alpha _t\ $are the temperature-renormalized parameters of
the anisotropy and interlayer coupling. We derive from for these quantities 
\begin{eqnarray}
f_t/f_r &=&\overline{\sigma }_r^{4/(N-1)}\left[ 1+{\cal O}(t_r/\overline{%
\sigma }_r^{1/\beta _2})\right]   \label{ft} \\
\alpha _t/\alpha _r &=&\overline{\sigma }_r^{2/(N-1)}\left[ 1+{\cal O}(t_r/%
\overline{\sigma }_r^{1/\beta _2})\right]   \label{at}
\end{eqnarray}
The leading logarithmic term in the square brackets of (\ref{MRAF})
corresponds to SSWT, while other two terms describe corrections to this
theory. As it should be expected, at low temperatures the RG results (\ref
{MRAF}), (\ref{ft}) and (\ref{at}) reduce to corresponding perturbation
expressions (\ref{MPAF}), (\ref{fr}) and (\ref{ar}). We have to bear in mind
that the quantity $t_r/\overline{\sigma }_r^{1/\beta _2}$ is a formal rather
than real estimate for neglected terms since higher order terms also yield a
contribution at not too low temperatures. Physically, neglecting such terms
is equivalent to neglecting 3D (or Ising-like) fluctuations in the RG
approach.

As already pointed, the Neel temperature cannot be calculated directly in
the RG approach since essentially non-spin-wave fluctuations contribute it
and an accounce of diagramms with an arbitrary number of loops is required.
However, one can obtain a general expression for the Neel temperature in the
following way. Consider first the temperature $t_r^{*}$ of the crossover to
the true critical region. This is determined by the condition $t_\rho \sim 1$%
, i.e. $t_r^{*}\sim \overline{\sigma }_r^{1/\beta _2}$ (the scale $\rho $ is
determined by (\ref{eq})). Substituting this into (\ref{MRAF}) one obtains 
\begin{equation}
t_r^{*}=2\left/ \left[ (N-2)\ln \frac 2{u_r^2\Delta (f^{*},\alpha ^{*})}%
+2\ln (2/t_r^{*})+C_{\text{AF}}\right] \right.
\end{equation}
where $f^{*}=f_r(t_r^{*})^{2/(N-2)},$ $\alpha ^{*}=\alpha
_r(t_r^{*})^{1/(N-2)},$ $C_{\text{AF}}$ is some constant of order of unity.
With further flow of RG parameters, 3D Heisenberg (or 2D Ising) fluctuations
change only the constant $C_{\text{AF}}$ which is replaced by the universal
function $\Phi _{\text{AF}}(\alpha _r/f_r)\sim 1.$ Thus for the Neel
temperature we have 
\begin{equation}
t_{\text{Neel}}=2\left/ \left[ (N-2)\ln \frac 2{u_r^2\Delta (f_c,\alpha _c)}%
+2\ln (2/t_{\text{Neel}})+\Phi _{\text{AF}}(\alpha _r/f_r)\right] \right.
\label{RGTN}
\end{equation}
where 
\begin{equation}
\,\,f_c=f_rt_{\text{Neel}}^{2/(N-2)},\,\,\alpha _c=\alpha _rt_{\text{Neel}%
}^{1/(N-2)}.
\end{equation}
[recall that $t=T/(2\pi \rho _s),$ $u_r=c/T$]. The function $\Phi $ is
determined by non-spin-wave fluctuations and cannot be calculated within the
RG approach. Note that the equation (\ref{RGTN}) is analogous to the result
for the correlation length in the renormalized classical regime in the 2D
case \cite{Chakraverty,Chubukov}, which may be rewritten in the same manner: 
\begin{equation}
t_r=2\left/ \left[ (N-2)\ln (\xi ^2/u_r^2)+2\ln (2/t_r)+\ln C_\xi \right]
\right.  \label{Ksi}
\end{equation}
where $C_\xi $ is an universal numerical preexponential factor. Comparing (%
\ref{RGTN}) and (\ref{Ksi}) we see that the only difference is that in the
quasi-2D anisotropic case the logarithms are cut at $\Delta ^{1/2}$ (rather
than at $1/\xi $ in the isotropic 2D case). In the quasi-2D isotropic case
the quantity $\Phi $ can be be calculated within the $1/N$ expansion (see
the next Section), and a more general case requires numerical analysis
(e.g., the quantum Monte-Carlo method).

Three limiting cases may be considered:

\noindent (i) anisotropic 2D case $\alpha =0.$ The equation for
magnetization (\ref{MRAF}) takes the form 
\begin{eqnarray}
\overline{\sigma }_r^{1/\beta _2} &=&1-\frac{t_r}2\left[ (N-2)\ln \frac 1{%
u_r^2f_r}+\frac 4{\beta _2}\ln (1/\overline{\sigma }_r)\right.  \label{MRAF1}
\\
&&\left. +2(1-\overline{\sigma }_r^{1/\beta _2})+{\cal O}(t_r/\overline{%
\sigma }_r^{1/\beta _2})\right]  \nonumber
\end{eqnarray}
The equation for the Neel temperature (\ref{RGTN}) reads 
\begin{equation}
t_{\text{Neel}}=2\left/ \left[ (N-2)\ln \frac 1{u_r^2f_r}+4\ln (2/t_{\text{%
Neel}})+\Phi _{\text{AF}}(\infty )\right] \right.  \label{TNAF1}
\end{equation}

\noindent (ii) isotropic quasi-2D case $f=0.$ We obtain 
\begin{eqnarray}
\overline{\sigma }_r^{1/\beta _2} &=&1-\frac{t_r}2\left[ (N-2)\ln \frac 2{%
u_r^2\alpha _r}+\frac 3{\beta _2}\ln (1/\overline{\sigma }_r)\right.
\label{MRAF2} \\
&&\left. +2(1-\overline{\sigma }_r^{1/\beta _2})+{\cal O}(t_r/\overline{%
\sigma }_r^{1/\beta _2})\right]  \nonumber
\end{eqnarray}
\begin{equation}
t_{\text{Neel}}=2\left/ \left[ (N-2)\ln \frac 2{u_r^2\alpha _r}+3\ln (2/t_{%
\text{Neel}})+\Phi _{\text{AF}}(0)\right] \right.  \label{TNAF2}
\end{equation}

\noindent (iii) large-$N$ case. To retain the structure of the equation (\ref
{eq}) at finite $N$, it is convenient to expand (\ref{eq})\ in $1/(N-2)$
rather than in $1/N.$ To first order we obtain 
\begin{eqnarray}
\rho ^2 &\simeq &\frac{\Delta (f_r,\alpha _r)}2\left[ 1+\frac{B_2-2}{N-2}\ln 
\frac{t_r}{t_\rho }\right]  \nonumber \\
&\simeq &\frac{\Delta (f_r,\alpha _r)}2\left( \frac{t_r}{t_\rho }\right)
^{(B_2-2)/(N-2)}
\end{eqnarray}
Then we have 
\begin{eqnarray}
\overline{\sigma }_r^{1/\beta _2} &=&1-\frac{t_r}2\left[ (N-2)\ln \frac 2{%
u_r^2\Delta (f_r,\alpha _r)}+B_2\ln (1/\overline{\sigma }_r^2)\right.
\label{MRAF3} \\
&&+2(1-\overline{\sigma }_r^2)+{\cal O}(t_r/\overline{\sigma }_r^2,1/N)\bigg]
\nonumber
\end{eqnarray}
and 
\begin{equation}
t_{\text{Neel}}=2\left/ \left[ (N-2)\ln \frac 2{u_r^2\Delta (f_r,\alpha _r)}%
+B_2\ln (2/t_{\text{Neel}})+\Phi _{\text{AF}}(\alpha _r/f_r)\right] \right.
\end{equation}
$\ $

The ferromagnetic case can be considered in a similar way. In this case the
quantum ground-state renormalizations are absent and the ``classical''
regime (an analogue of the ``renormalized classical'' one in the
antiferromagnetic case) is determined by 
\begin{equation}
T\gg JS\max \{f,\alpha \}  \label{cond1}
\end{equation}
To develop perturbation theory for the partition function (\ref{zf}) we pass
from the real fields $\pi _x,\pi _y$ to the cyclic components 
\begin{equation}
\pi ^{\pm }=\pi _x\pm i\pi _y
\end{equation}
and expand again square roots in $\pi ^{+},\pi ^{-}.$ The bare Green's
function of the fields $\pi ^{+},\pi ^{-}$ takes the form 
\begin{equation}
G^{(0)}({\bf p},i\omega _n)=\frac 1g\left[ i\omega _n+p_{\parallel
}^2+\alpha (1-\cos p_z)+f+h\right] ^{-1}
\end{equation}
Due to absence of quantum renormalizations we have $Z_{Qi}\equiv 1$ and the
indices $r$ may be dropped$.$ The factors $\widetilde{Z}_i$ has the same
form (\ref{RAF}) as in the antiferromagnetic case with $N=3$ and the
replacement $\ln (u_r\mu )\rightarrow \ln (u\mu )/2$.

The relative magnetization $\overline{\sigma }\equiv \overline{S}/S$ ($%
\overline{S}=\langle S^z\rangle $) to the two-loop approximation reads 
\begin{eqnarray}
\overline{\sigma } &=&1-\frac t2\ln \frac 2{u\Delta (f,\alpha )}  \nonumber
\\
&&-\frac{t^2}4(B_2-2)\ln \frac 2{u\Delta (f,\alpha )}
\end{eqnarray}
($t=g/u$). The scaling equation (\ref{mrg}) is valid in the ferromagnetic
case too. The equation for $\rho $\ in the this case takes the form 
\begin{equation}
2\rho ^2=u\Delta (f_\rho ,\alpha _\rho )
\end{equation}
Thus we obtain the equation for the magnetization in the form 
\begin{eqnarray}
\overline{\sigma } &=&1-\frac t2\left[ \ln \frac 2{u\Delta (f_t,\alpha _t)}%
+2\ln (1/\overline{\sigma })\right.  \label{MRF} \\
&&+2(1-\overline{\sigma })+{\cal O}(t/\overline{\sigma })\bigg]  \nonumber
\end{eqnarray}
The results for the temperature renormalization of the anisotropy and
interlayer coupling parameters have the same form (\ref{ft}), (\ref{at}).
The Curie temperature is determined in the same way as the Neel temperature
in the antiferromagnetic case. The result reads 
\begin{equation}
t_{\text{Curie}}=2\left/ \left[ \ln \frac 2{u\Delta (f_c,\alpha _c)}+2\ln
(2/t_{\text{Curie}})+\Phi _{\text{F}}(\alpha /f)\right] \right.  \label{TC}
\end{equation}

The calculation of the magnetization and ordering temperature of a classical
magnet is performed in Appendix B.

Thus the RG approach is sufficient to calculate the magnetization in the
spin-wave and the 2D-like regions and to calculate the Neel\ (Curie)
temperature up to some universal constant. The crossover temperature region
of the quantum antiferromagnet can be considered within the $1/N$ expansion.
Besides that, in the case of quantum quasi-2D antiferromagnet, this
expnasion enables one to describe the true critical region and to evaluate
the quantity $\Phi _{\text{AF}}(0).$

\section{Comparison with the $1/N$ expansion in the quantum $O(N)$ model and
the crossover to the critical regime}

The $1/N$ expansion gives a possibility to develop another perturbation
theory for the partition function (\ref{zaf}). Unlike the renormalization
group approach, this method works satisfactorily at arbitrary temperatures.
As well as in Section 3, we consider only the ordered phase.

Consider first the case of the antiferromagnet. We begin with the
generalization of the results of Ref.\cite{Quasi2D} to the case where
anisotropy is present. To develop perturbation theory in $1/N$ we integrate
out $\sigma $-fields from (\ref{zaf}). Thus we have 
\begin{eqnarray}
{\cal Z}_{AF}\, &=&\int D\lambda \exp (NS_{eff}[\lambda ,h])  \label{Zef} \\
S_{eff}[\lambda ,h] &=&\frac 12\ln \det \widehat{G}_0+\frac 1{2g}(1-%
\overline{\sigma }^2)\text{Sp}(i\lambda )  \nonumber \\
&&+\frac 1{2g}\text{Sp}\left[ \left( i\lambda \overline{\sigma }-h/\rho
_s^0\right) \widehat{G}_0\left( i\lambda \overline{\sigma }-h/\rho
_s^0\right) \right]  \label{Sef}
\end{eqnarray}
where 
\begin{eqnarray}
\widehat{G}_0^{mm} &=&[\partial _\tau ^2+{\bf \nabla }^2+\alpha \Delta
_z+f(1-\delta _{mN})]^{-1} \\
\Delta _z\sigma _{i_z}({\bf r,}\tau {\bf )} &=&\sigma _{i_z+1}({\bf r,}\tau 
{\bf )}-\sigma _{i_z}({\bf r,}\tau {\bf )}  \nonumber
\end{eqnarray}
and $\overline{\sigma }=\langle \sigma _N({\bf r},\tau )\rangle $ is the
relative staggered magnetization.

Since $N$ enters (\ref{Zef}) only as a prefactor in the exponent, expanding
near the saddle point generates a series in $1/N.$ To zeroth order in $1/N$
the excitation spectrum, which is given by the poles of the unperturbed
longitudinal and transverse Green's functions, contains a gap $f^{1/2}$ for
all the components $\sigma _m$ except for $m=N$: 
\begin{eqnarray}
G_t^0({\bf q},\omega _n) &=&\left[ \omega _n^2+q_{\Vert }^2+\alpha (1-\cos
q_z)+f\right] ^{-1} \\
\,\,\,G_l^0({\bf q},\omega _n) &=&\left[ \omega _n^2+q_{\Vert }^2+\alpha
(1-\cos q_z)\right] ^{-1}.
\end{eqnarray}
The absence of the gap for the $N$-th (longitudinal) mode in the ordered
phase is an {\it exact }property of the model under consideration in any
order in $1/N.$

The sublattice magnetization at $T<T_{\text{Neel}}$ is determined by the
constraint equation $\langle \sigma ^2\rangle =1.$ To first order in $1/N$
the constraint takes the form 
\begin{eqnarray}
1-\overline{\sigma }^2 &=&gT\frac{N-1}N\sum_{\omega _m}\int \frac{d^2{\bf k}%
_{\Vert }}{(2\pi )^2}\int\limits_{-\pi }^\pi \frac{dk_z}{2\pi }G_t^0({\bf k}%
,\omega _m)  \nonumber \\
&&\ \ \ \ \ \ \ \ +g\left[ F(T,\overline{\sigma })-R(T,\overline{\sigma }%
)\right] ,  \label{Constr1/Nm}
\end{eqnarray}
where 
\begin{eqnarray}
R(T,\overline{\sigma }) &=&T\sum_{\omega _m}\int \frac{d^2{\bf k}_{\Vert }}{%
(2\pi )^2}\int\limits_{-\pi }^\pi \frac{dk_z}{2\pi }\left[ G_t^0({\bf k}%
,\omega _m)\right] ^2  \label{R} \\
&&\ \ \ \ \ \ \times \left[ \Sigma _t({\bf k},\omega _m)-\Sigma _l({\bf 0}%
,0)\right] ,  \nonumber \\
F(T,\overline{\sigma }) &=&\frac{2T}N\sum_{\omega _m}\int \frac{d^2{\bf k}%
_{\Vert }}{(2\pi )^2}\int\limits_{-\pi }^\pi \frac{dk_z}{2\pi }G_l^0({\bf k}%
,\omega _m)\frac{\Pi ({\bf k},\omega _m)}{\widetilde{\Pi }({\bf k},\omega _m)%
}  \label{F}
\end{eqnarray}
The longitudinal and transverse mode self-energies are given by 
\begin{equation}
\Sigma _{t,l}({\bf k},\omega _m)=\frac{2T}N\sum_{\omega _n}\int \frac{d^2%
{\bf q}_{\Vert }}{(2\pi )^2}\int\limits_{-\pi }^\pi \frac{dq_z}{2\pi }\frac{%
G_{t,l}^0({\bf k-q},\omega _m-\omega _n)}{\widetilde{\Pi }({\bf q},\omega _n)%
}
\end{equation}
where 
\begin{eqnarray}
\widetilde{\Pi }({\bf q},\omega _n) &=&\Pi ({\bf q},\omega _n)+\frac{2%
\overline{\sigma }^2}gG_l^0({\bf q},\omega _n), \\
\Pi ({\bf q},\omega _n) &=&T\sum_{\omega _l}\int \frac{d^2{\bf p}_{\Vert }}{%
(2\pi )^2}\int\limits_{-\pi }^\pi \frac{dp_z}{2\pi }  \label{Pi} \\
&&\ \times \left[ \frac{N-1}NG_t^0({\bf p},\omega _l)G_t^0({\bf q}+{\bf p}%
,\omega _n+\omega _l)\right.  \nonumber \\
&&\ \left. +\frac 1NG_l^0({\bf p},\omega _l)G_l^0({\bf q}+{\bf p},\omega
_n+\omega _l)\right] .  \nonumber
\end{eqnarray}
Since the polarization operator $\Pi (q,\omega _n)$ enters only the
first-order corrections in (\ref{Constr1/Nm}), the contribution from
longitudinal Green's functions to $\Pi $ influences thermodynamic quantities
in order of $1/N^2$ and can be formally neglected. However, one should bear
in mind that for finite $N$ this corrections may be large. Physically, the
neglection of the longitudinal part of $\Pi $ corresponds to neglecting the
contribution of Ising-like (spin-flip) excitations. However, these
excitations are negligible only at wavevectors $q\gg f^{1/2}$ (see, e.g.,
Ref.\cite{Landau}). As follows from (\ref{Pi}), such quasimomenta yield a
dominant contribution at $T\ll 2\overline{\sigma }^2[(\alpha +f)/f]^{1/2}/g.$
The opposite case $T\gg 2\overline{\sigma }^2[(\alpha +f)/f]^{1/2}/g$
corresponds to the Ising critical region which cannot be treated within the $%
1/N$ expansion.

As well as in the previous Section, we consider only the case $T\gg (\max
\{f,\alpha \})^{1/2}c.$ The procedure of integration and frequency summation
in (\ref{Constr1/Nm}) is analogous to that of Refs. \cite{Chubukov,Quasi2D}.
We obtain 
\begin{eqnarray}
&&1-\frac T{4\pi \rho _s}\left[ (N-2)\ln \frac{2T^2}{c^2\Delta }+B_2\ln 
\frac{\ln (T^2/c^2\Delta )+x_{\overline{\sigma }}}{x_{\overline{\sigma }}}%
\right.  \nonumber \\
&&\left. -2\frac{\ln (2T^2/c^2\Delta )}{\ln (2T^2/c^2\Delta )+x_{\overline{%
\sigma }}}-I_1(x_{\overline{\sigma }})\right]  \nonumber \\
\ &=&\overline{\sigma }_r^2\left[ 1+\frac 1N\ln \frac{\ln (2T^2/c^2\Delta
)+x_{\overline{\sigma }}}{x_{\overline{\sigma }}}-I_2(x_{\overline{\sigma }%
})\right]  \label{ConstrFF}
\end{eqnarray}
where $\Delta \equiv \Delta (f_r,\alpha _r),\,\,$(see Eq. (\ref{Delta})), $%
B_2$ is determined by (\ref{b2}), 
\begin{equation}
x_{\overline{\sigma }}=\frac{4\pi \rho _s}{(N-2)T}\overline{\sigma }_r^2
\end{equation}
and we have introduced the quantum-renormalized parameters 
\begin{equation}
f_r=f(1-2Q_\Lambda ),\,\,\alpha _r=\alpha (1-Q_\Lambda )  \label{rrs}
\end{equation}
\begin{equation}
\,\rho _s=(1+4Q_\Lambda )\rho _s^{N=\infty },\,\,\,\overline{\sigma }%
_r^2=g\rho _s(1-Q_\Lambda )/N  \label{rs}
\end{equation}
with $\rho _s^{N=\infty }=Nc(1/g-\Lambda /2\pi ^2)$ is the renormalized spin
stiffness in zeroth order in $1/N,$ $Q_\Lambda =(8/3\pi ^2N)\ln (N\Lambda
c/16\rho _s)$ (in this Section we use the relativistic cutoff $\omega
_n^2+k^2<\Lambda ^2$ of frequency summations and quasimomentum integrations;
for this regularization scheme the bare spin wave velocity $c_0$ is replaced
by the quantum-renormalized one, $c$). Since another regularization scheme
is used, the expressions (\ref{rrs}), (\ref{rs}) are different from the
corresponding results of Sect.3. As well as in Sect.3, the
quantum-renormalized parameters are not universal and therefore should be
determined from the spin-wave theory (see Appendix A) rather than from the
continual model. The functions $I_{1,2}(x)$ are some functions with the
asymptotics $1/x$ at large $x,$ so that at $x_{\overline{\sigma }}\gg 1$
their contributions are small. For the isotropic quasi-2D case these
functions were calculated in Ref.\cite{Quasi2D}.

Consider first the case of not too high temperatures 
\begin{equation}
T(N-2)/4\pi \rho _s<\overline{\sigma }_r^2,  \label{reg1}
\end{equation}
where $I_{1,2}(x_{\overline{\sigma }})$ are small enough. Using the identity 
$\ln (T^2/c^2\Delta )+x_{\overline{\sigma }}=4\pi \rho _s/(N-2)T$ which
holds to zeroth order in $1/N,$ we transform the logarithmic term in the
right-hand side of (\ref{ConstrFF}) into a power to obtain 
\begin{eqnarray}
&&\left[ 1-I_2(x_{\overline{\sigma }})\right] \overline{\sigma }_r^{1/\beta
_2} 
\begin{array}{c}
=
\end{array}
1-\frac T{4\pi \rho _s}  \nonumber \\
&&\times \left[ (N-2)\ln \frac{2T^2}{c^2\Delta }+B_2\ln (1/\overline{\sigma }%
_r^2)-2+2\overline{\sigma }_r^2-I_1(x_{\overline{\sigma }})\right]
\label{Constr2D}
\end{eqnarray}
Note that in the quasi-2D case ($f=0$) the equation (\ref{Constr2D})
slightly differs from the equation (54) of Ref. \cite{Quasi2D} by the
replacement 
\[
\ln (2T^2/\alpha _r)/[\ln (2T^2/\alpha _r)+x_{\overline{\sigma }%
}]\rightarrow 1-\overline{\sigma }_r^2 
\]
which holds to zeroth order in $1/N$.

At temperatures $T(N-2)/4\pi \rho _s\ll \overline{\sigma }_r^2$ the
contribution of $I_1,I_2$ can be neglected and the fluctuations in this
temperature region have a 2D-like Heisenberg nature. In particular, in the
low-temperature region $T(N-2)\ln (2T^2/\Delta )/4\pi \rho _s\ll \overline{%
\sigma }_r^2$ the result of SSWT \cite{Our1st,OurFMM,OurSSWT} 
\begin{equation}
\overline{\sigma }_r=1-\frac{T(N-1)}{8\pi \rho _s}\ln \frac{2T^2}{c^2\Delta }
\end{equation}
is reproduced. One can see also that at $I_1=I_2=0$ the result (\ref
{Constr2D})\ coincides with the large-$N$ limit of the RG result (\ref{MRAF}%
) (see Eq. (\ref{MRAF3})). However, at finite $N$ the renormalization group
provides a more correct description of the sublattice magnetization at $%
T(N-2)/4\pi \rho _s\ll \overline{\sigma }_r^2$.

In the temperature region $\overline{\sigma }_r^2\sim T(N-2)/4\pi \rho _s,$
which corresponds to the crossover to the true critical behavior, the
situation changes. In this case the large-$N$ result demonstrates a more
sharp decrease of $\overline{\sigma }$ than the RG approach. In the quasi-2D
case \cite{Quasi2D} the result (\ref{Constr2D}) is smoothly joined with the
3D temperature dependence (see below). Thus in the quasi-2D case the result
of the $1/N$ expansion should be considered as an interpolation between the
low-temperature and critical regions. One could expect that this holds also
in the presence of the anisotropy were critical behavior cannot be described
within the $1/N$ expansion.

In the region $x_{\overline{\sigma }}\ll 1,$ i.e. $\overline{\sigma }_r^2\ll
T(N-2)/4\pi \rho _s$ the true critical behavior takes place. In the
isotropic quasi-2D case ($f=0$) the result for the staggered magnetization
reads (see Ref. \cite{Quasi2D}) 
\begin{equation}
\overline{\sigma }_r^2=\left[ \frac{4\pi \rho _s}{(N-2)T_{\text{Neel}}}%
\right] ^{\beta _3/\beta _2-1}\left[ \frac 1{1-A_0}\left( 1-\frac T{T_{\text{%
Neel}}}\right) \right] ^{2\beta _3}  \label{MagnCr}
\end{equation}
where $\beta _3=\left( 1-8/\pi ^2N\right) /2$ is the true 3D critical
exponent for the order parameter (for $N=3$ we have $\beta _3\simeq 0.36$)$%
,\,A_0=2.8906/N$. The result of the $1/N$ expansion for $T_{\text{Neel}}$ 
\cite{Quasi2D} satisfies the general formula (\ref{TNAF2}) with 
\begin{equation}
\Phi _{\text{AF}}(0)=-0.0660  \label{TN1}
\end{equation}

In the anisotropic 2D case ($\alpha =0$) the temperature dependence of
magnetization in the critical region is determined by the Ising-like
excitations (domain walls) which cannot be considered within the $1/N$
expansion. The universality hypothesis predicts the same temperature
behavior of $\overline{\sigma }$ as in Ising systems 
\begin{equation}
\overline{\sigma }_r^8=A(1-T/T_{\text{Neel}}),  \label{i}
\end{equation}
where $A$ is some constant. As demonstrate numerical calculations (see the
next Section), the temperature dependence of $\overline{\sigma }$ determined
from (\ref{Constr2D}) is smoothly joined with (\ref{i}), $A$ and $T_{\text{%
Neel}}$ being considered as fitting parameters.

In the presence of both anisotropy and interlayer coupling, the situation in
the critical region $x_{\overline{\sigma }}<1$ is more complicated$.$
Consider first the case $f<\alpha .$ Then at $1>x_{\overline{\sigma }%
}>[f/(\alpha +f)]^{1/2}$ the (sublattice) magnetization has the 3D
Heisenberg behavior (\ref{MagnCr}) with some coefficient $1/[1-A_0(f/\alpha
)]$; at $x_{\overline{\sigma }}<[f/(\alpha +f)]^{1/2}$ the behavior $%
\overline{\sigma }(T)$ changes to the 3D Ising one. At $f\sim \alpha $ the
3D Heisenberg region disappears, and in the whole critical region the 3D
Ising behavior takes place. With further increase of $f$ (at $f>\alpha $),
the 2D Ising critical region occurs for $1>x_{\overline{\sigma }%
}>x_0(f/\alpha )$, while at $x_{\overline{\sigma }}<x_0(f/\alpha )$ the 3D
Ising behavior still takes place. However, the dependence $x_0(f/\alpha )$
cannot be calculated within the approaches under consideration.

Now we turn to the case of a ferromagnet. As already mentioned, in this case
the dynamical part of the action cannot be generalized to arbitrary $N.$
Thus the expressions (\ref{Constr1/Nm})-(\ref{Pi})\ of the first order in $%
1/N$ (where we put $N=3$) should be considered as physically reasonable
rather than strict results. The dynamical part of the action in (\ref{zf})
results in cutting the quasimomentum integrals at $q\sim (T/JS)^{1/2}.$ The
indices $r$ can be dropped, since the quantum renormalizations are absent.
Thus we have instead of (\ref{Constr2D}) for $T\gg JS\max \{f,\alpha \}$%
\begin{equation}
\overline{\sigma }=1-\frac T{4\pi \rho _s^0}\left[ \ln \frac{2T}{JS\Delta }%
+B_2\ln (1/\overline{\sigma }^2)-2+2\overline{\sigma }^2+{\cal O}(4\pi \rho
_s^0/\overline{\sigma }^2)\right]  \label{1/NF}
\end{equation}
where ${\cal O}(4\pi \rho _s^0/\overline{\sigma }^2)$ terms cannot be
calculated within such a consideration. For the Curie temperature we
reproduce the RG result (\ref{TC}).

\section{Discussion and comparison with experimental data}

The above consideration provides a description of the long-range order of
quantum and classical magnets in different temperature regions. Let us
summarize the results obtained in the practically interesting case $N=3$. In
the spin-wave and 2D-like regions, i.e. at 
\begin{equation}
\overline{\sigma }_r\gg T/4\pi \rho _s,\,\Gamma \gg \Delta  \label{regs}
\end{equation}
we have the RG result for the relative (sublattice) magnetization 
\begin{equation}
\overline{\sigma }_r=1-\frac T{4\pi \rho _s}\left[ \ln \frac{2\Gamma (T)}{%
\Delta (f_t,\alpha _t)}+2\ln (1/\overline{\sigma }_r)+2(1-\overline{\sigma }%
_r)\right]  \label{fRG}
\end{equation}
where the function $\Delta (f,\alpha )$ is determined by (\ref{Delta}), the
temperature-renormalized values of interlayer coupling and anisotropy
parameters are 
\begin{equation}
f_t/f_r=(\alpha _t/\alpha _r)^2=\overline{\sigma }_r^2  \label{fat}
\end{equation}
and the quantities $\Gamma (T),\overline{\sigma }_r,f_r,\alpha _r,\rho _s$
are given in the Table 1 (see also Appendix A). 
\[
\begin{tabular}{||llllll||}
\hline\hline
& $\Gamma (T)$ & $\overline{\sigma }_r$ & $\rho _s$ & $f_r$ & $\alpha _r$ \\ 
$\text{quantum }AFM$ & $T^2/c^2$ & $\overline{S}/\overline{S}_0$ & $\gamma S%
\overline{S}_0$ & $f\overline{S}_0^2/S^2$ & $\alpha \overline{S}_0/S$ \\ 
$\text{quantum }FM$ & $T/JS$ & $\overline{S}/S$ & $\rho _s^0$ & $f$ & $%
\alpha $ \\ 
$\text{classical }FM,AFM$ & $32$ & $\overline{S}/S$ & $\rho _s^0Z_{L1}$ & $%
fZ_{L2}^{-1}$ & $\alpha Z_{L3}^{-1}$ \\ \hline\hline
\end{tabular}
\]
Corresponding equation for the magnetic ordering temperature $T_M$ has the
form 
\begin{equation}
T_M=4\pi \rho _s\left/ \left[ \ln \frac{2\Gamma (T_M)}{\Delta (f_c,\alpha _c)%
}+2\ln \frac{4\pi \rho _s}{T_M}+\Phi (f/\alpha )\right] \right.  \label{tTc}
\end{equation}
where $\Phi (x)$ is some function of order of unity (in the quantum case it
is universal, i.e. does not depend on the upper cutoff parameter), $f_c$ and 
$\alpha _c$ are the temperature-renormalized interlayer coupling and
anisotropy parameters at $T=T_M$ that are determined by 
\begin{equation}
f_c/f_r=(\alpha _c/\alpha _r)^2=(T_M/4\pi \rho _s)^2
\end{equation}
Since $T_M/4\pi \rho _s\sim 1/\ln (1/\Delta )\ll 1$ the temperature
renormalizations are important when treating experimental data. In
particular, the parameters, which are measured at different temperatures,
may differ considerably.

In the case $\alpha =0$ we have 
\begin{eqnarray}
\overline{\sigma }_r &=&1-\frac T{4\pi \rho _s}\left[ \ln \frac{\Gamma (T)}{%
f_r}+4\ln (1/\overline{\sigma }_r)+2(1-\overline{\sigma }_r)\right]
\label{fp1} \\
T_M &=&4\pi \rho _s\left/ \left[ \ln \frac{\Gamma (T_M)}{f_r}+4\ln \frac{%
4\pi \rho _s}{T_M}+\Phi (0)\right] \right.  \label{fp2}
\end{eqnarray}
In the case $f=0$ we obtain 
\begin{eqnarray}
\overline{\sigma }_r &=&1-\frac T{4\pi \rho _s}\left[ \ln \frac{2\Gamma (T)}{%
\alpha _r}+3\ln (1/\overline{\sigma }_r)+2(1-\overline{\sigma }_r)\right]
\label{fp3} \\
T_M &=&4\pi \rho _s\left/ \left[ \ln \frac{2\Gamma (T_M)}{\alpha _r}+3\ln 
\frac{4\pi \rho _s}{T_M}+\Phi (\infty )\right] \right.  \label{fp4}
\end{eqnarray}
The results of solving the SSWT equations\cite{OurFMM,OurSSWT} differ from (%
\ref{fp1})-(\ref{fp4}) by the replacement $4(3)\rightarrow 2(1)$ for the
coefficient at the second term in the square brackets (which yields the
double-logarithmic correction to the standard SWT) in the anisotropic 2D
(isotropic quasi-2D) case respectively. Thus the role of the corrections to
SSWT is more important in the isotropic quasi-2D case than in the 2D
anisotropic one.

As discussed in the Introduction, the results (\ref{fRG}), (\ref{tTc}) are
expected to hold in the first order of the $1/M$ expansion in the $CP^{M-1}$
model (at $M=2$).

In the temperature interval outside the critical region 
\begin{equation}
\overline{\sigma }_r^2>T/4\pi \rho _s,\,\Gamma \gg \Delta  \label{int}
\end{equation}
the result of the $1/N$ expansion in the $O(N)$ model to the first order in $%
1/N$ reads 
\begin{eqnarray}
&&\ \ \ \left[ 1-I_2(x_{\overline{\sigma }})\right] \overline{\sigma }_r 
\begin{array}{c}
=
\end{array}
1-\frac T{4\pi \rho _s}\left[ \ln \frac{2\Gamma (T)}{\Delta (f_r,\alpha _r)}%
\right.  \nonumber \\
&&\ \ \ \left. +2B_2\ln (1/\overline{\sigma }_r)+2(1-\overline{\sigma }%
_r^2)+I_1(x_{\overline{\sigma }})\right]  \label{f1/N}
\end{eqnarray}
where $x_{\overline{\sigma }}=4\pi \rho _s\overline{\sigma }_r^2/T,$ $B_2$
and $\Delta $ are determined by (\ref{b2}),\ (\ref{Delta}), $I_{1,2}(x)$ are
some functions with the asymptotics $1/x$ at large $x,$ other quantities are
given in the Table 1. In particular cases $\alpha =0$ and $f=0$ the
coefficient at the second term in the square brackets in (\ref{f1/N}) is too
times larger than for the RG results (\ref{fp1}), (\ref{fp3}). In the
spin-wave and 2D-like temperature regions this is an artifact of the
first-order $1/N$ expansion. At the same time, the $1/N$ expansion provides
a more correct description of the crossover temperature region. Due to the
difference in the crossover conditions (\ref{int}) and (\ref{regs}), the
equations for $T_M$ have the same form (\ref{fp2}), (\ref{fp4}) in both
approaches.

Now we discuss the experimental situation. First we consider the temperature
dependence of the sublattice magnetization in La$_2$CuO$_4$ (Ref. \cite
{Keimer}) which is shown in Fig.1. This Figure presents also the results of
spin-wave approximations (SWT, SSWT and the Tyablikov theory\cite{Tyab}, see
a more detailed discussion in Ref.\cite{Quasi2D}), RG approach and the
result of $1/N$ expansion (\ref{f1/N}). The value $\gamma \simeq 1850K$ was
calculated by using the experimental data\cite{Veloc} while $\alpha
_r=1\cdot 10^{-3}$ was obtained from the best fit of experimental dependence 
$\overline{\sigma }_r(T)$ to the spin-wave theory at low temperatures. The
result of the $1/N$ expansion to first order in $1/N$ is $T_{\text{Neel}%
}=345 $ K which is considerably lower than for all the spin-wave
approximations and is in a good agreement with the experimental value, $T_{%
\text{Neel}}^{\exp }=325$K. The RG approach describes correctly the
dependence $\overline{\sigma }_r(T)$ in the spin-wave region $(T<300K)$ and
2D-like region (which is very narrow since $\alpha $ is very small) while at
higher temperatures this approach overestimates $\overline{\sigma }.$ At the
same time, the $1/N$ expansion curve is most close to the experimental data
and demonstrates a correct critical behavior. The results of the numerical
solution of equation (\ref{f1/N}) in the temperature region (\ref{int}) and
the dependence (\ref{MagnCr}) in the critical region turn out to be smoothly
joined at the point $T=330$K (marked by a cross).

In the crossover region ($320$K$<T<340$K) the theoretical $O(3)$ curve lies
slightly higher than the experimental one. One may speculate that this is
due to the influence of anisotropy. Fixing $\Delta $ in (\ref{f1/N}) and
determining $B_2$ from the best fit at intermediate temperatures (see Fig.1)
one finds the values $\alpha _r=1\cdot 10^{-4},\,\,\,f_r=5\cdot 10^{-4}.$
This value of $\alpha $ is more close to the experimental data of Ref.\cite
{ic}. Thus our approach gives a possibilty to estimate the relative role of
interlayer coupling and magnetic anisotropy in layered compounds.

In the layered perovskites K$_2$NiF$_4,$ Rb$_2$NiF$_4$ and K$_2$MnF$_4$ the
magnetic anisotropy is known to be more important than the interlayer
coupling. K$_2$NiF$_4$ has spin $S=1,$ and neutron scattering data yield $%
|J|=102$K and $T_{\text{Neel}}^{\exp }=97.1$K (see Ref.\cite{Joungh}). Fig.2
shows the experimental dependence $\overline{\sigma }(T)$ \cite{Birgeneau}
and the results of the spin-wave approaches, RG approach and the numerical
solution of Eq.(\ref{f1/N}). The value $f_r=0.0088$ was obtained from the
best fit of the result of SSWT to experimental data at low temperatures
(this value coincides well with the experimental one $f_r=0.0084,$ Ref. \cite
{Joungh}). In the spin-wave and 2D-like temperature intervals (\ref{regs}) ($%
T<80$K) the curves corresponding to the $1/N$ expansion and RG approach lie
somewhat higher than the experimental points since $T^2/f_rc^2$ in this
region is not large, and the renormalized-classical description is not too
good (a more accurate calculation can be performed by carrying out exact
summation over the Matsubara frequencies). Bearing in mind this correction,
the RG approach gives a more correct qualitative tendency than the $1/N$%
-expansion in the 2D-like region. At the same time, the $1/N$ expansion
curve is in a good numerical agreement with experimental data. The joining
procedure with the Ising critical behavior (\ref{i}) may be performed in a
rather wide temperature region $0.85T_{\text{Neel}}<T<0.9T_{\text{Neel}}$
and gives $A=0.01,$ $T_{\text{Neel}}=91.4$K. The width of the critical
``Ising'' region makes up about $1$K. Note that an account of the terms of
order of $1/x_{\overline{\sigma }}$ in (\ref{Constr2D}), which can be
performed by analogy with the calculations of Ref.\cite{Quasi2D}, gives $T_{%
\text{Neel}}=92.7$K.

In the crossover region ($80$K$<T<90$K) the theoretical $O(3)$ curve for K$%
_2 $NiF$_4$ lies, in contrast with the case of La$_2$CuO$_4,$ slightly lower
than the experimental one. This fact may be attributed to the influence of
interlayer coupling. The fitting in the crossover region yields the values $%
\alpha _r=0.0017,\,\,\,f_r=0.0069$ which correspond to $T_{\text{Neel}}=97$K
and the bare parameters $\alpha \,|J\,|=0.1\,$K$,\,\,\zeta |J|=0.76$K.
Direct experimental data for $\alpha $ are absent, but our estimation seems
to be reasonable.

Rb$_2$NiF$_4$ has a larger magnetic anisotropy. According to Ref.\cite
{Joungh}, one has $|J\,|=82$K, $|J|f_r=3.45$K, $T_{\text{Neel}}^{\exp }=94.5$%
K. From the best fit of SSWT to the dependence $\overline{\sigma }_r(T)$ at
low temperatures one obtains $f_r=0.046$ which is also in a good agreement
with the above experimental value$.$ Then one obtains from (\ref{f1/N}) $T_{%
\text{Neel}}=95.5$K.

K$_2$MnF$_4$ has spin $S=5/2$ and therefore represents a situation which is
intermediate between the quantum and classical cases. Fig.3 shows a
comparison of the results of different approaches with experimental data for
this compound. The parameters used are $|J|=8.4$K, $|J|f_r=0.13$K (see Ref.%
\cite{Joungh}). One can see that the $1/N$ expansion yields good results,
and the experimental points lie between the quantum and classical RG curves,
the quantum approximation being essentially more satisfactory. This confirms
once more that it is difficult to realize the classical limit (see Appendix
B). Note that SSWT, which correctly takes into account lattice effects,
provides in this case better results in comparison with the RG approach.
Thus an accurate treatment of such situations within continual models
requires numerical calculations of quasimomentum integrals and sums over
Matsubara frequencies in (\ref{Constr1/Nm}).

Fig.4 shows a comparison of the results of SSWT and the RG approach for the
magnetization of a classical magnet with the Monte-Carlo calculations \cite
{Levanjuk}. One can see that, except for a very narrow critical region, the
RG curve is rather accurate although topological excitations are neglected.
Note that the region of applicability of the RG approach in the classical
case is more broad than in the quantum case, so that we need not to use the
large-$N$ approach for describing the crossover to the critical region.

Thus the RG approach (or, equivalently, the $1/M$ expansion in the $SU(M)$
model) and the $1/N$ expansion in the $O(N)$ model turn out to give good
results in different temperature regions. Whereas the first method describes
well the 2D-like regime, the $1/N$ expansion describes successfully the
crossover to the critical region and, in the absence of anisotropy, the
critical behavior. Both methods give the same results for the ordering point
up to the terms of order of $\ln \ln (1/\Delta )$. Besides that, the $1/N$
expansion permits to calculate nonsingular terms in the quasi-2D case.

To conclude, our results give a possibility to describe magnetic properties
of real layered magnets with a rather high accuracy. The approaches applied
may be useful for treating magnetic and structural phase transitions in
systems with more complicated order parameters\cite{Baxter}.

\section*{Appendix A. Spin-wave results for the ground-state
renormalizations in a quantum antiferromagnet.}

The ground-state thermodynamic quantities of the quantum antiferromagnet can
be calculated within the spin-wave theory. The result for the ground-state
staggered magnetization reads (see, e.g., \cite{ArovasBook}) 
\begin{equation}
\overline{S}_0=S-\frac 12\sum_{{\bf k}}\left[ \frac 1{\sqrt{1-\phi _{{\bf k}%
}^2}}-1\right] \simeq S-0.1966  \label{GSS}
\end{equation}
where $\phi _{{\bf k}}=\frac 12(\cos k_x+\cos k_y).$ The ground-state spin
stiffness and spin-wave velocity to first order in $1/S$ are given by \cite
{ArovasBook} 
\begin{equation}
\rho _s=\gamma S\overline{S}_0,\,\,\,\,c=\sqrt{8}\gamma S  \label{GSR}
\end{equation}
with $\gamma $ being the quantum-renormalized intralayer exchange parameter
determined by 
\begin{equation}
\gamma \,/\,|\,J\,|=1+\frac 1{2S}\sum_{{\bf k}}\left[ 1-\sqrt{1-\phi _{{\bf k%
}}^2}\right] \approx 1+\frac{0.0790}S  \label{GSG}
\end{equation}
For the quantum-renormalized coupling constant we have $g_r=c/\rho _s$. The
quantum-renormalized interlayer coupling and anisotropy parameters can be
determined from the first-order $1/S$-corrections to the excitation
spectrum. We have at $T=0$ and for small in-plane wavevector components \cite
{OurSSWT} 
\begin{equation}
E_{{\bf k}}^2\simeq 8(\gamma S)^2\left[ k_{\Vert }^2+\frac{2\gamma ^{\prime }%
}\gamma (1-\cos k_z)+\frac \delta {\gamma S}\right]  \label{ESW}
\end{equation}
where 
\begin{equation}
\gamma ^{\prime }=\frac \alpha 2(\overline{S}_0/S)|\,J\,|
\end{equation}
is the renormalized interlayer coupling and 
\begin{equation}
\delta =(\overline{S}_0/S)^2\left[ (2S-1)\zeta +4\eta S\,|\,J\,|\,/\gamma
\right] \,|\,J\,|
\end{equation}
is the renormalized anisotropy. Note that in the case $\zeta ,\eta \ll 1,$
which is considered only, single- and two-site anisotropies lead to the same
effects. Comparing the spectrum (\ref{ESW}) with the bare spin-wave spectrum
determined from (\ref{G0AF}) 
\begin{equation}
E_{{\bf k}}^2=c^2\left[ k_{\Vert }^2+\alpha (1-\cos k_z)+f\right]
\end{equation}
we obtain the relation between the bare and the quantum-renormalized
parameters: 
\begin{eqnarray}
f_r &=&\frac \delta {\gamma S}=(\overline{S}_0/S)^2\left[ \frac{(2S-1)\zeta
\,|\,J\,|}{\gamma S}\,+\frac{4\eta J^2}{\gamma ^2}\right] \\
\alpha _r &=&\frac{2\gamma ^{\prime }}\gamma =\alpha \overline{S}_0/S
\end{eqnarray}
However, it should be noted that since the spectrum (\ref{ESW}) contains
only renormalized parameters rather than the bare ones, $(\alpha /2)|J|$ and 
$[2\zeta (1-1/2S)+4\eta ]\,|\,J|,$ only $\gamma ^{\prime }$ and $\delta $
can be determined experimentally.

\section*{Appendix B. Renormalization group analysis in the lattice $O(N)$
model and the limit of classical spins}

The treatment of the partition function for the classical anisotropic
quasi-2D anisotropic magnets (\ref{zcl}) is similar to the isotropic 2D case%
\cite{Chakraverty}. In this case the relative temperature $t=T/(2\pi \rho
_s^0)$ plays the role of a coupling constant, and we have instead of the
first line of (\ref{ZR}) the scaling relation 
\begin{equation}
t=Z_1t_R
\end{equation}
where $t_R$ is the renormalized temperature. The bare Green's function of
the field $\mbox {\boldmath $\pi $}={\bf n-(nz)z}$ has the form 
\begin{eqnarray}
G^{(0)}({\bf q}) &=&\frac 1{2\pi t}[2(2-\cos q_xa-\cos q_ya) \\
&&+\alpha (1-\cos q_za)+f+h]^{-1}
\end{eqnarray}
where $a$ is the lattice constant. The renormalization constants can be
calculated from the two-point vertex function. It is useful to represented
this constants as 
\begin{equation}
Z_i(t,a)=Z_{Li}(t)\widetilde{Z}_i(t_L,a)
\end{equation}
where $t_L=tZ_{L1}^{-1},$ $Z_{Li}$ contain non-logarithmic terms which are
not changed under RG transformations, and $\widetilde{Z}_i$ contain all the
other terms. We have 
\begin{eqnarray}
Z_{L1} &=&Z_{L2}=Z_{L3}=1-\pi t/2+{\cal O}(t^2)  \nonumber \\
Z_L &=&1
\end{eqnarray}
The results for $\widetilde{Z}_i$ read 
\begin{eqnarray}
\widetilde{Z} &=&1+t_L(N-1)\ln (64a\mu )  \nonumber \\
&&\ +t_L^2(N-1)(N-3/2)\ln ^2(64a\mu )+{\cal O}(t_L^3), \\
\widetilde{Z}_1 &=&1+t_L(N-2)\ln (64a\mu )+t_L^2(N-2)\ln ^2(64a\mu )+{\cal O}%
(t_L^3),\,  \nonumber \\
\widetilde{Z}_2 &=&1-2t_L\ln (64a\mu )+{\cal O}(t_L^2),\,\,\,\,\widetilde{Z}%
_3=1-t_L\ln (64a\mu )+{\cal O}(t_L^2),  \label{RCL}
\end{eqnarray}
The expression for the magnetization to two-loop approximation is 
\begin{eqnarray*}
\overline{\sigma } &=&1-\frac{t_L(N-1)}4\ln \frac{64}{\Delta (f_L,\alpha _L)}
\\
&&\ +\frac{t_L^2(3-N)(N-1)}{32}\ln ^2\frac{64}{\Delta (f_L,\alpha _L)} \\
&&\ -\frac{t_L^2(N-1)}8\left[ 1+\frac{f_L}{\sqrt{f_L^2+2\alpha _Lf_L}}%
\right] \ln \frac{64}{\Delta (f_L,\alpha _L)}
\end{eqnarray*}
where we have defined the renormalized quantities in the lattice case, 
\begin{equation}
\,f_L=fZ_{L2}^{-1},\,\,\alpha _L=\alpha Z_{L3}^{-1}
\end{equation}
The equation for the magnetization can be derived in the same way as in
Sect.2 to obtain 
\begin{eqnarray}
\overline{\sigma }^{1/\beta _2} &=&1-\frac{t_L}2\left[ (N-2)\ln \frac{64}{%
\Delta (f_t,\alpha _t)}+\frac 2{\beta _2}\ln (1/\overline{\sigma })\right.
\label{MRCL} \\
&&\ \ \ \ \ \ \ \ \ \ \ \ \ \left. +2(1-\overline{\sigma }^{1/\beta _2})+%
{\cal O}(t_L/\overline{\sigma }^{1/\beta _2})\right]  \nonumber
\end{eqnarray}
Then the ordering temperature satisfies the equation 
\begin{equation}
t_M=2\left/ \left[ (N-2)\ln \frac{64}{\Delta (f_c,\alpha _c)}+2\ln
(1/t_M)+\Phi _{\text{cl}}(\alpha /f)\right] \right.  \label{TCC}
\end{equation}
Note that in the case $\alpha =0$ a similar expression was obtained in Ref.%
\cite{RGA}. However, the result of this work contains wrong coefficient at
the second term in the square brackets of (\ref{TCC}) since not all two-loop
corrections were taken into account.

Comparing the result for the magnetization (\ref{MRCL}) with these of
Sect.2, (\ref{MRAF}) and (\ref{MRF}), we can write down the criteria of
applicability of the classical limit (see Sect.2) with the correct numerical
factors 
\begin{eqnarray}
T^2 &\gg &32c^2,\;\;\text{AFM,}  \label{crit} \\
T &\gg &32JS,\;\;\text{FM.}  \nonumber
\end{eqnarray}
It is difficult to satisfy these criteria in the ordered phase $T<T_M\sim
1/\ln (1/\Delta )$ at not too small $\Delta $ and $1/S$ due to the large
value of the numerical factor in (\ref{crit}).

{\sc Figure captions}

Fig.1. The theoretical temperature dependences of the relative sublattice
magnetization $\overline{\sigma }_r$ from different spin-wave
approximations, RG approach (Eq. (\ref{fp3})) and $1/N$ expansion in the $%
O(N)$ model (Eqs. (\ref{Constr2D}) and (\ref{MagnCr})), and the experimental
points for La$_2$CuO$_4$ (Ref.\cite{Keimer}). The RG curve is shown up to
the temperature where the derivative $\partial \overline{\sigma }_r/\partial
T\ $diverges. The curve denoted by $1/N^{\prime }$ is the best fit in the
crossover temperature region to the experimental data with the anisotropy
being the fitting parameter (see discussion in the text).

Fig.2. The relative staggered magnetization $\overline{\sigma }_r(T)$ for K$%
_2$NiF$_4$ (points) as compared to the standard spin-wave theory
(long-dashed line), SSWT (dot-dashed line), RG approach and result of the
solution of (\ref{Constr2D}) in the intermediate-temperature region (\ref
{int}) (solid line). The short-dashed line shows the extrapolation of the $%
1/N$-expansion result to the Ising-like critical region according to Eq.(\ref
{i}). The boundary of the 2D-like and crossover regions is marked by an
arrow.

Fig.3. The dependence $\overline{\sigma }_r(T)$ for K$_2$MnF$_4$ (points) as
compared to the results of SSWT (dashed line), RG analysis in the quantum
(dot-dot-dashed line) and classical (dot-dashed line) limits and solution of
(\ref{Constr2D}) (solid line).

Fig.4. The renormalization group (solid line) and SSWT (dashed line) results
for the relative magnetization $\overline{\sigma }$ of a classical
anisotropic 2D magnet ($\zeta =0,$ $\eta =0.001$) as compared to the result
of the Monte-Carlo calculation\cite{Levanjuk}. The RG and SSWT curves are
shown up to the temperature where $\partial \overline{\sigma }/\partial
T=\infty .$

{\sc Table caption}

Table 1. Parameters of the equations for the (sublattice) magnetization (\ref
{fRG}), (\ref{f1/N}) for different cases, $Z_{L1}=Z_{L2}=Z_{L3}=1-T/8\pi
\rho _s^0$.

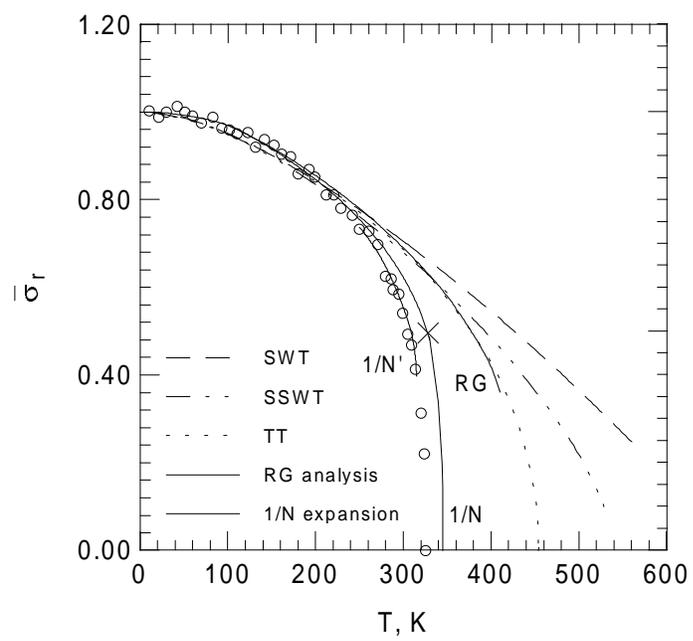
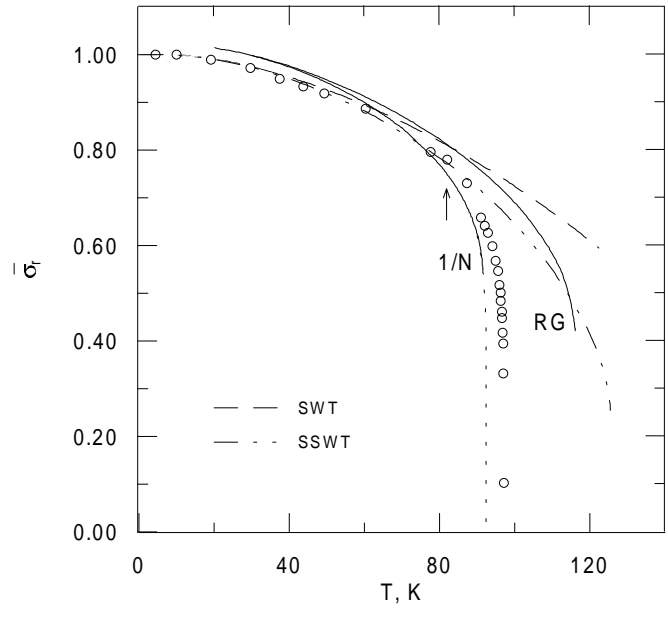
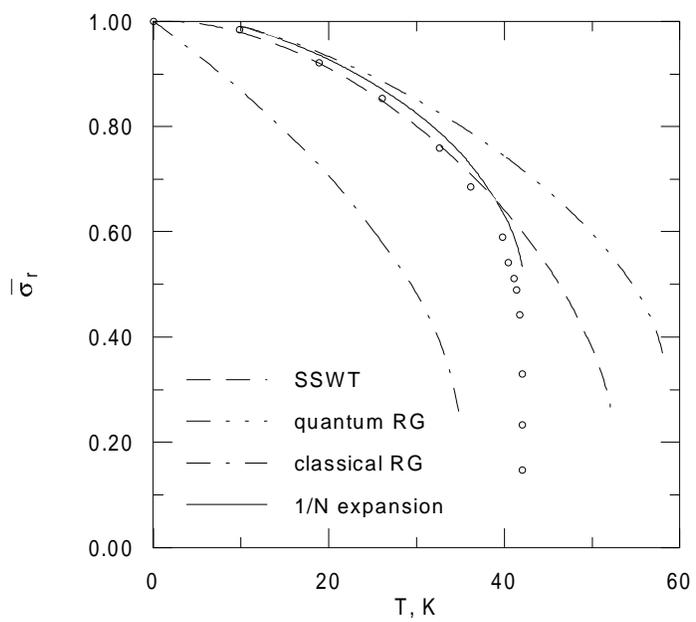
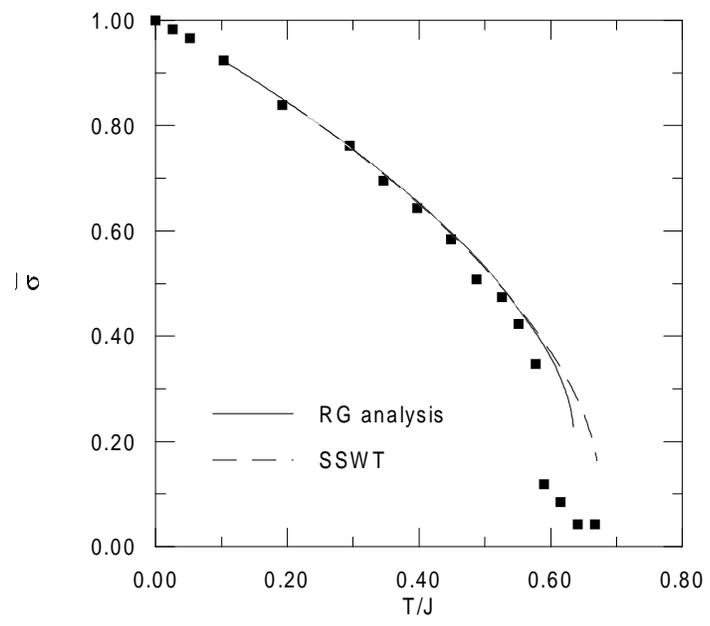
\end{document}